\input{aipcheck}

\documentclass[
    ,final            
  ]
  {aipproc}

\layoutstyle{6x9}

\newcommand{\sol}{$_{\odot}$}

\begin{document}

\title{Probing Stellar Populations at z $\sim$ 7 -- 8}


\author{Steven L. Finkelstein}{
  address={Texas A\&M University\\ George P. and Cynthia Woods Mitchell Institute for Fundamental Physics and Astronomy\\ Department of Physics and Astronomy\\4242 TAMU, College Station, TX 77845\\ stevenf@physics.tamu.edu}
}

\begin{abstract}
In this proceeding we present the results from a study of very high--redshift galaxies with the newly commissioned Wide Field Camera 3 on the {\it Hubble Space Telescope}.  With the deepest near-infrared data ever taken, we discovered 31 galaxies at 6.3 $<$ z $\leq$ 8.6.  The rest--frame ultraviolet (UV) colors of these galaxies are extremely blue, showing significant ($>$ 4 $\sigma$) evolution from z $\sim$ 3, over only 1 Gyr of cosmic time.  While we cannot yet diagnose the exact cause of the bluer colors, it appears a low dust content is the primary factor.  The stellar masses of these galaxies are less than comparably selected galaxies at 3 $<$ z $<$ 6, highlighting evolution in the stellar mass of characteristic (L$^{\ast}$) galaxies with redshift.  Lastly, the measured rest--UV luminosity density of galaxies in our sample seems sufficient to sustain reionization at z $\sim$ 7 when we account for the likely contribution from galaxies below our magnitude limit.
\end{abstract}

\maketitle


\section{Introduction}

Until recently, the robust identification of galaxies at z $\geq$ 7, corresponding to a time when the Universe was only $\sim$ 800 Myr old, was extremely difficult due to existing near--infrared instrumentation.  The recent installation of the Wide Field Camera 3 (WFC3) on the {\it Hubble Space Telescope} ({\it HST}) has opened up the distant universe for detailed exploration for the first time.  WFC3 has obtained the deepest near--infrared (NIR) images ever taken, in the {\it Hubble} Ultra Deep Field (HUDF), obtaining data in three filters, Y$_{105}$, J$_{125}$, and H$_{160}$ (PID 11563, PI G. Illingworth).  In conjunction with existing optical Advanced Camera for Surveys (ACS) data, these new data allowed selection of galaxies at z $>$ 6.5 by searching for the Lyman break, due to intervening intergalactic medium (IGM) H\,{\sc i} absorption.  This is typically done by choosing a set of color criteria which select galaxies at the desired redshift, while excluding interlopers such as low--redshift passively evolving galaxies and galactic brown dwarfs, which can mimic high--redshift galaxies in a single color.  Soon after these data were released, samples of z $>$ 6.5 galaxies were published, mostly using color cuts \cite[e.g.,][]{bouwens10, oesch10, mclure10, bunker09, yan09}.  In these proceedings, we present the results of an independent study, using photometric redshift techniques to select a high--redshift galaxy sample, as well as rest--frame ultraviolet (UV) color analysis and stellar population modeling to study the physical characteristics of these galaxies.  This full study is published in \cite{finkelstein10}.

\section{High--Redshift Galaxy Sample}
We fit photometric redshifts to objects in the WFC3 data using the photometric redshift code EAZY \cite{brammer08}.  This takes into account all available data, including the ACS photometry, when computing the redshift, and computes the full redshift probability distribution function, which is a sum over all available templates.  This also provides a better redshift estimate, $\Delta$z $\pm$ 0.15--0.3, versus $\Delta$z $\pm$ 0.5 for the Lyman break technique.  We found 31 objects which met our three selection criteria: 1) $>$ 3.5 $\sigma$ significance in the J$_{125}$ {\it and} H$_{160}$ bands; 2) Best--fit photometric redshift of 6.3 $<$ z$_\mathrm{phot}$ $\leq$ 8.6; and 3) $\geq$ 60\% of the integrated probability distribution function ($\mathcal{P}_6$ ) at z $>$ 6.  

\begin{figure}[!h]
\includegraphics[height=.28\textheight]{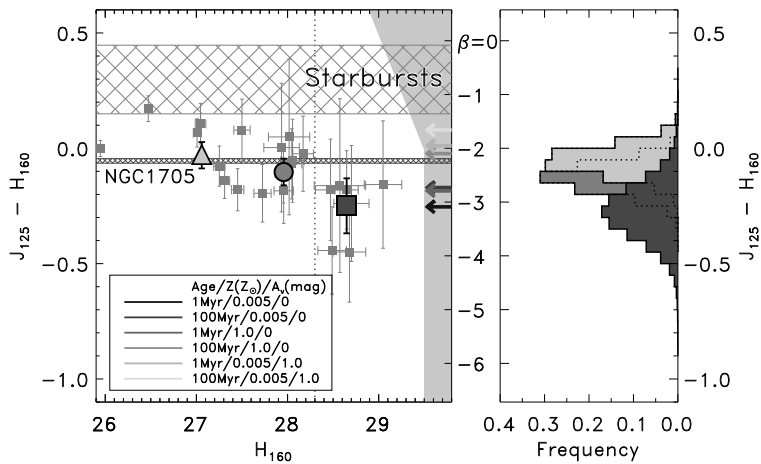}
\caption{$H$ versus $J - H$ diagram for the $z \sim$ 7 sample, plotted as gray squares.  The circle, triangle and large square denote the mean color for all objects, the bright subsample (H $<$ 28.3) and the faint subsample (H $\geq$ 28.3), respectively.  The histograms show the distribution of the means for these samples from our simulations, where the spread in these distributions provides the errors on the sample means.  We also show the colors spanned by local starbursts, NGC~1705, as well as a number of stellar popualtion models.  The mean of all objects is bluer than local starbursts at $>$ 4 $\sigma$ significance, exhibiting evolution from z $\sim$ 3, where galaxies have colors consistent with local starbursts.  The faint objects have a mean bluer than NGC~1705 by $<$ 2 $\sigma$, thus we cannot yet robustly probe the existence of exotic stellar populations.}
\end{figure}

\section{Rest--Frame UV Colors of z $\sim$ 7 Galaxies}
Figure 1 shows the 23 objects in our z $\sim$ 7 sub--sample (6.3 $<$ z$_\mathrm{phot}$ $<$ 7.5) on a color--magnitude diagram.  Individual galaxies have blue rest--UV colors, $-$0.5 $<$ $J-H$ $<$ 0.2, corresponding to a UV spectral slope of $-$4 $<$ $\beta$ $<$ $-1$.  However, the faint nature of these galaxies results in a high uncertainty on their colors.  Thus, to learn about the z $\sim$ 7 galaxy population as a whole, we computed the mean color of all objects in the z $\sim$ 7 sample, as well as the mean color in bright and faint bins (split at $H$ = 28.3).  We ran 10$^{7}$ bootstrap Monte Carlo simulations to assess the uncertainty on these mean colors, where each simulation accounts for Poisson noise in the sample size as well as photometric scatter.  The means and their corresopnding uncertainties are shown in Figure 1.  We compare these means to the colors of local starburst galaxies \cite{kinney96}, as well as the local extremely blue galaxy NGC~1705 (which is thought to be dust--free), and find that z $\sim$ 7 galaxies are bluer than local starbursts at $>$ 4 $\sigma$ significance (i.e., 99.9987\% confidence).  This is in stark contrast to the properties of Lyman break galaxies at z $\sim$ 3, which, only a billion years later, have rest--frame UV colors consistent with local starbursts \cite[i.e.,][]{papovich01, reddy09, bouwens09}.  Thus there is significant evolution in the rest--frame UV properties of star--forming galaxies over only $\sim$ 1 Gyr of cosmic time.  The faint galaxies in our z $\sim$ 7 sample have $\beta$ = $-$3.07, which is difficult for ``normal'' stellar populations to form, thus suggesting the presence of a top--heavy initial mass function (IMF), or zero metallicity (Population III) stars \cite[e.g.,][]{bouwens10b}.  However, the photometric scatter at H $>$ 28.3 results in a large uncertainty of $\sigma_\beta$ = $\pm$ 0.5.  Thus, while exotic stellar populations {\it could} be present, these data do not provide strong evidence to support their existence.

\begin{figure}[h]
\includegraphics[height=.28\textheight]{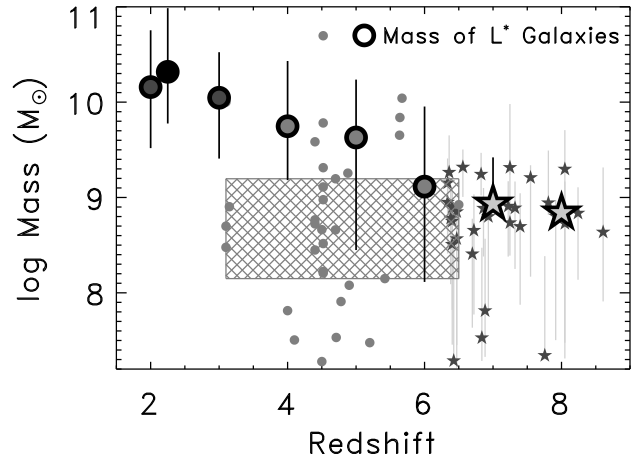}
\caption{The stellar masses of L$^{\ast}$ LBGs versus redshift (dark gray--\cite{reddy06}; black--\cite{shapley05}; gray--\cite{stark09} at z = 4, 5 and 6).  The mass at L$^{\ast}$ at $z = $7--8 from our results are shown by the large stars.  The error bars represent the central 80\% range of the data.  We show our individual galaxies as small stars with their uncertainties, which are typically a factor of +2/$-$4.5.  The background gray circles denote stellar masses of Ly$\alpha$~emitting galaxies at 3.1 $\leq$ z $\leq$ 6.5 from the literature \cite[e.g.,][]{gawiser06b,pirzkal07,finkelstein07,finkelstein09a}.  The gray hatched region denotes the interquartile range of the LAE masses.  The masses of the  z $>$ 6.3 LBGs studied here are more similar to those of LAEs at all redshifts than LBGs at any redshift $<$ 6.}
\end{figure}

\section{Stellar Mass Evolution}
We compared the spectral energy distributions (SEDs) of our sample to updated stellar population models of S.\ Charlot \& G.\ Bruzual \cite{bruzual03}.  Unfortunately, with only NIR detections (i.e. upper limits in the optical and in the infrared with {\it Spitzer}) we are unable to robustly constrain the ages and dust properties of these galaxies, though their very blue colors imply low levels of dust extinction.  However, these blue colors do place limits on the metallicities, with a 68\% confidence of Z $<$ 0.05 Z\sol.  Even though these galaxies are not detected in their rest--frame optical (observed infrared) we place constraints on their stellar masses as the young age of the Universe limits the amount of mass which can exist in old stars, unseen in the WFC3 bands.  We find best--fit masses in the range of $\sim$ 10$^{8}$ -- 10$^{9}$ M\sol.  Figure 2 shows the masses of our galaxies (interpolated to L$^{\ast}$ using the luminosity functions of \cite{oesch10} and \cite{bouwens10}) compared to the stellar masses of lower--redshift LBGs (also converted to the mass at L$^{\ast}$).  Comparing the masses of L$^{\ast}$ galaxies is roughly equivalent to studying them at a constant number density, thus we trace the stellar masses of the direct progenitors and descendents of a given galaxy.  Figure 2 also shows the masses of galaxies selected on the basis of their Ly$\alpha$ emission (LAEs), which are typically less evolved than LBGs \cite[i.e.,][]{gawiser06b, finkelstein09a}.  The masses of z $\geq$ 7 galaxies are more comparable to LAEs at all redshifts than LBGs at any lower redshift.  The low masses and likely low--dust content of z $\geq$ 7 galaxies implies that they are physically similar to lower--redshift LAEs rather than scaled--down versions of LBGs.  Lower redshift LAEs are the likely building blocks of more evolved galaxies \cite[i.e.,][]{gawiser07}, thus at z $\sim$ 7 it appears we are observing an epoch dominated by these building blocks, and that the more evolved galaxies common at lower--redshifts are rare.  

\section{Summary}
We have analyzed the properties of 31 galaxies at z $>$ 6.3 in the extremely deep HUDF.  These objects appear very similar to lower--redshift LAEs, implying that at z $\geq$ 7 we have reached the era of ``baby'' galaxies.  However, it does not appear we have yet reached the era of ``infant'' galaxies, where one would expect Population III stars or top heavy IMFs, but these may come to light in the near future with the {\it James Webb Space Telescope}.


\begin{theacknowledgments}
I would like to thank my collaborators on this work, Casey Papovich, Mauro Giavalisco, Naveen Reddy, Harry Ferguson, Anton Koekemoer and Mark Dickinson.
\end{theacknowledgments}





\IfFileExists{\jobname.bbl}{}
 {\typeout{}
  \typeout{******************************************}
  \typeout{** Please run "bibtex \jobname" to optain}
  \typeout{** the bibliography and then re-run LaTeX}
  \typeout{** twice to fix the references!}
  \typeout{******************************************}
  \typeout{}
}

\end{document}